\documentclass[]{elsart}

\usepackage{epsfig}
\usepackage{amssymb}
\begin{document}

\begin{frontmatter}
\title{Interaction-round-a-face density-matrix renormalization-group
method}
\author[wada]{Wada Tatsuaki\corauthref{cor}}
\corauth[cor]{Corresponding author.}and
\ead{wada@ee.ibaraki.ac.jp} 
\author[nishino]{Nishino Tomotoshi}
\address[wada]{Department of Electrical and Electric Engineering, 
Faculty of Engineering, Ibaraki University, Hitachi 316-8511, Japan}

\address[nishino]{Department of Physics, Faculty of Science, 
Kobe University, 657-8501, Japan}

\date{\today}
\maketitle

\begin{abstract}
We demonstrate the numerical superiority of the interaction-round-a-face (IRF)
density-matrix renormalization-group (DMRG) method applied to $SU(2)$
invariant quantum spin chains over the conventional DMRG.
The ground state energy densities and the gap energies of both $S = 1$ and
$S = 2$ spin chains can be calculated using the IRF-DMRG without extensive
computations. We have also studied the effect of tuning boundary interaction
$J_{\rm end}^{~}$ at both ends of the chain from the IRF view point.
It is clearly observed that the magnon distribution is uniform when
the best $J_{\rm end}^{~}$ is chosen.
\end{abstract}

\end{frontmatter}

\section{introduction}

Quantum spin chains are one of the most extensively studied systems
by way of numerical calculations. In this article we consider isotropic
spin-$S$ anti-ferromagnetic Heisenberg (AFH) spin chain, whose
Hamiltonian is represented as
\begin{equation}
   H = J \sum_{i=1}^{N-1} {\bf S}_i \cdot {\bf S}_{i+1},
   \label{H}
\end{equation}
where $J$ is a positive coupling constant and ${\bf S}_i^{~}$ denotes
spin operator at $i$-th site. After Haldane conjectured the
existence of finite excitation gap for integer spin chains and the absence
for half-integer ones \cite{Hal83}, many efforts have been performed to
confirm this conjecture. We focus on numerical analysis of the AFH
model in the following.

A simple but concrete way to obtain the gap energy of the
spin-$S$ AFH chain in the thermodynamic limit is to numerically calculate
the excitation energies for finite size systems, and to perform the finite size
scaling. Conventionally Lanczos method is applied to the Hamiltonian 
(Eq.~(\ref{H}))
expressed as a large sparse matrix in order to obtain lower lying states and 
their energies.
The major shortcoming of this way is that the Lanczos diagonalization
requires huge numerical resources. This is because the matrix dimension
of the $N$-site system is of the order of $(2S+1)^N$, that rapidly increases
with both $N$ and $S$. The largest system size $N$ which is manageable 
with current computer is around 36 (for $S=1/2$) even now.

The Density-matrix renormalization-group (DMRG) invented by
White~\cite{Ref-1,Ref-2} is a novel and powerful computational
method, which overcomes the limitation of the manageable
system size in the Lanczos
diagonalization. It enables us to obtain physical observables such as
ground state energy, correlation functions, etc., of large scale
systems very accurately without performing extensive computations.
It is not overwhelming to say that the DMRG is one of the standard
tools to study low dimensional quantum systems.
DMRG is a real-space renormalization method, which divides the whole system
into the left and the right blocks, and reduces the degree of freedom
of each block using the eigenvalues of the reduced density
matrix (DM) as probabilistic measures to choose the relevant block-spin
states. The number of the block spin state kept under the RG
transformation is conventionally represented by the letter $m$.
It is important to choose sufficiently large $m$ in order to keep
numerical precision. The computational cost in DMRG is, roughly speaking,
proportional to $m^3_{~}$, which is not heavily dependent on size $N$.

When applying DMRG to the spin-$S$ AFH chain
specified by Eq. (\ref{H}), the block spin states have total spin $L$ up 
to $nS$, 
where $n$ is a non-negative integer and depends on the number of spins 
in the block, we have to keep sufficient numbers of the block-spin state
up to a certain total-spin $L$. This is
problematic for large $S$ chains, because the system has spin
rotational symmetry; the eigenvalues of the DM are $(2L+1)$-fold
degenerated, since a block spin state with spin $L$ belongs to $2L+1$
spin multiplet. Consequently the necessary block spin states
($= m$) increases rapidly with $S$.

The problem of $(2L+1)$-fold degeneracy can be overcome by integrating out
the $SU(2)$ symmetry of the spin chain, and by representing the AFH
Hamiltonian using the
total spin basis. As a result, the AFH system can be represented as a
quantum limit of the {\it interaction round a face} (IRF) model~\cite{Bax82}.
Sierra and Nishino \cite{Ref-3} have modified DMRG algorithm for
the $S = 1/2$ and $1$ AFH models in the IRF representation (rep), where their
numerical procedure is called `IRF-DMRG'. Wada~\cite{Ref-4} has further 
developed
IRF-DMRG for the cases $S = 1$ and $S = 2$.
It is easily understood that
the the computational cost of IRF-DMRG is much lower than that of the
conventional DMRG, since the dimension of the associated Hamiltonian matrix
in IRF rep is much smaller than that in conventional ($=$ vertex) rep.

\section{IRF-DMRG applied to $S = 1$ and $2$ AFH systems}

Now we demonstrate the numerical efficiency of IRF-DMRG when it is
applied to AFH model for higher spin ($S \ge 1$) chains. In order to
treat large $S$ systems, Wada~\cite{Ref-4} has developed a way of 
obtaining the IRF
expression for the Heisenberg interaction for $S \ge 1$.
In the IRF formulation, the spin excited states can be targeted by
putting an additional auxiliary (or ghost) spin to the spin chain;
more details are shown in Ref.~\cite{Ref-4}.

Before showing the calculated results we stress that all the numerical data are
obtained using personal computer with the moderate power
(SPECfp95 $\sim 25$). This is because the Hilbert space dimension in the
IRF rep is much smaller than that in the
conventional $S^z_{~}$ basis rep. For example, in the case of
spin-$2$ spin chain, to keep $m_{\rm IRF} \sim 90$ states in the IRF rep
is equivalent to keep $m_{s_z} \sim 450$ in the conventional rep. We
keep 100 states at most, in the IRF rep.
The ground state energy density $e_0$ and the estimated gap energy $\Delta$
for $S = 1$ are $e_0 = -1.401484039 J$ and $\Delta = 0.4104 J$, and those
for $S = 2$ are $e_0 = -4.76124816 J$ and $\Delta = 0.088 J$~\cite{Ref-4}.

Having described the computational benefits of IRF formulation, now let us
observe the boundary effect from the IRF view point. Since DMRG (including
the IRF-DMRG) chiefly treats systems with open boundary, it is important
to decrease the boundary effect and obtain the properties of bulk state.

\begin{figure}
\begin{center}
  \epsfig{file=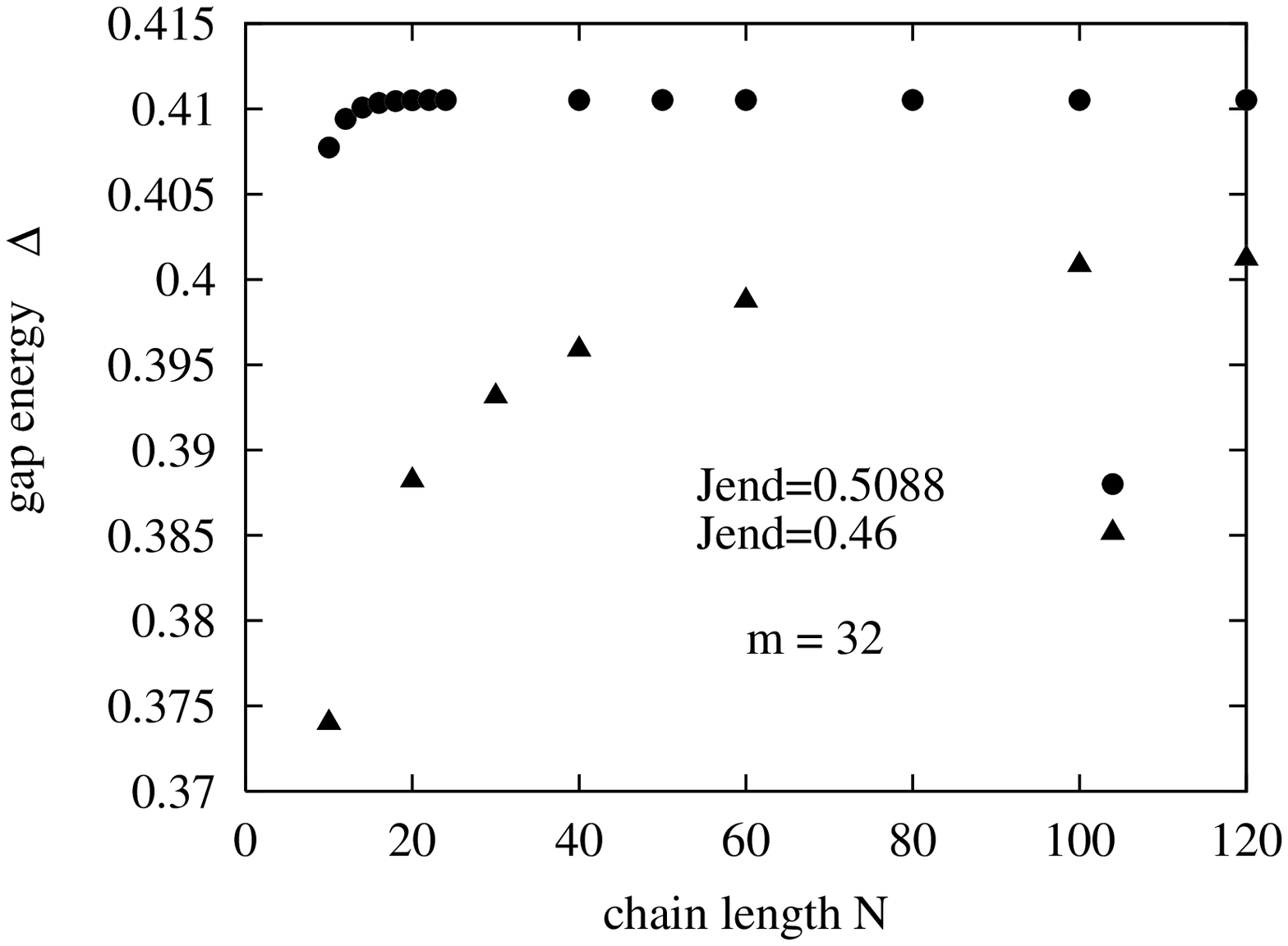,width=125mm}
\end{center}
\caption{The gap energies $\Delta$ as a function of the chain length $N$
for spin-$1$ AFH spin chain. The number of states $m$ being kept during IRF-
DMRG iterations are only 32. The $J_{\rm end}$ denotes the coupling constants
at both ends of the chain. All coupling constants $J$ except $J_{\rm end}$ 
set to
unity. When $J_{\rm end}$ sets to $0.5088$, $\Delta(N)$ is
almost independent of $N$.
}
\label{fig:Egap}

\end{figure}

In order to estimate the spin excitation gap of AFH in the thermodynamic
limit, White {\it et~al.}~\cite{Whi93} minimized the boundary effect by
tuning the coupling constants $J_{\rm end}$ at both ends of the spin chain.
They found that for $S = 1$ chain the best $J_{\rm end}$ is $0.5088$ for
the first excited state, and is $0.7$ for the ground state. Let us review
how fast the boundary effect decreases with the system size when
$J_{\rm end} = 0.5088$. Figure~\ref{fig:Egap} shows the gap energies
$\Delta$ as a function of the system size $N$. After rapid increase with $N$,
the gap energy $\Delta (N)$ becomes almost independent of $N$
when it is larger than $20$. It should be noted that it is possible to
estimate $\Delta=0.4015 J$ from the data $N \le 20$.
In contrast, when $J_{\rm end}$ deviates from the best value, $\Delta(N)$
shows very slow increases with $N$, this makes the finite size scaling
very difficult.

\begin{figure}
\begin{center}
  \epsfig{file=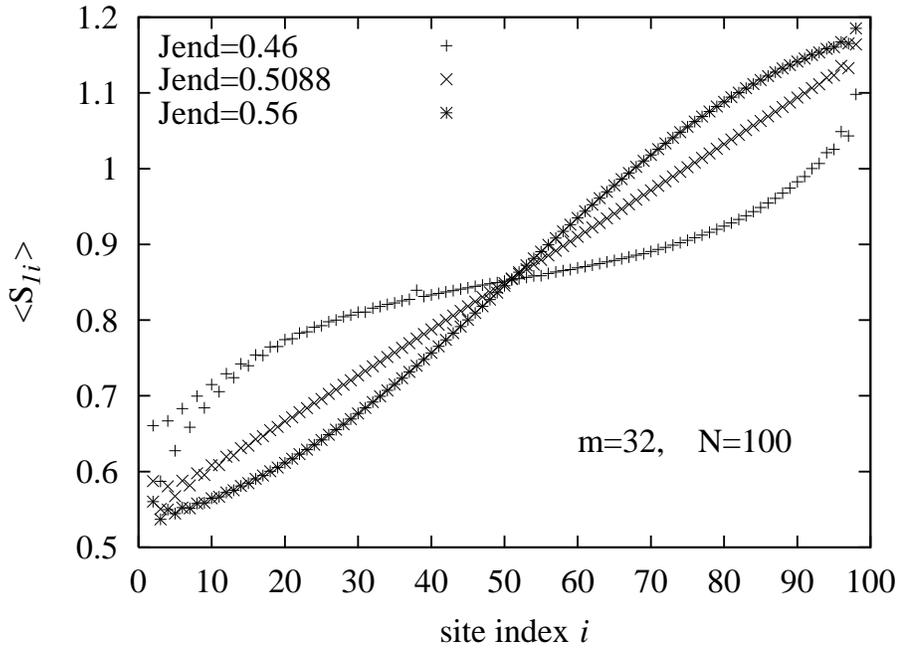,width=125mm}
\end{center}
\caption{The partial spin average $\langle S_{1i} \rangle$ distribution for
the first excited state of the spin-$1$ AFH chain.  When $J_{\rm end}=0.5088$,
at which the boundary effect is minimized, $\langle S_{1i} \rangle$ 
becomes linear
in site index $i$.}
\label{fig:Save}
\end{figure}

What happens when $J_{\rm end}=0.5088$ for the first excited state?
It is believed that the magnon (or the domain wall excitation)  has
zero momentum ($k = \pi$ in their definition) and it distributes quite
uniformly~\cite{Whi93}. The situation is quite similar to the case of a
particle in tight binding model with shallow negative potential at the
system boundary.
The IRF rep detects the magnon distribution more clearly,
since it is possible to obtain the partial spin average
\begin{equation}
\langle S_{1i}^{~} \rangle \equiv \langle \sum_{j=1}^{i} S_j^{~} \rangle \, .
\end{equation}
The derivative of $\langle S_{1i}^{~} \rangle$ with respect to the lattice
coordinate $i$ is proportional to the density of the quasi particle
($=$ magnon).
To speak precisely, it is possible to obtain $\langle S_{1i}^{~} \rangle$
from the numerical data of the conventional DMRG, but careful treatment
of numerical data is required; the IRF-DMRG provides $\langle S_{1i}^{~} 
\rangle$
more directly. Figure \ref{fig:Save} shows $\langle S_{1i}^{~} \rangle$ for
three different $J_{\rm end}\;$s. As is seen the partial spin average is quite
linear in $i$ at the best boundary condition (BC) of $J_{\rm end} = 0.5088$. 
The partial spin average
shall be a good reference data when searching the similar BC
for the higher spin chains. More detailed analysis will be published elsewhere.

In conclusion we have demonstrated the numerical advantage of IRF formulation
through the study of the AFH model, particularly for higher spin chains.
The ground state energy densities $e_0$ and gap energies $\Delta$ are
obtained with moderate computational cost. The results are
$e_0 = -1.401484039 J$, $\Delta = 0.4104 J$ for $S = 1$,
and $e_0 = -4.76124816 J$, $\Delta = 0.088 J$ for $S = 2$.
The estimated spin gap is much more accurate than
those obtained so far, under the condition $m \le 100$.
In addition, we have discussed how the minimization of the boundary effect
is observed from the IRF view point. We have seen that at the best boundary
condition at which $J_{\rm end}$ minimizes the boundary effect, the partial
spin average $\langle S_{1i}^{~} \rangle \equiv \langle \sum_{j=1}^{i}
S_j^{~} \rangle$ of the first excited state is almost linear in $i$.
To find out such smooth BCs for higher spin cases is
our future subject.

T.~N thank to G.~Sierra for valuable discussions. The present work is supported
by Grant-in-Aid for  Scientific Research from
Ministry of Education, Science, Sports and Culture
(No.~11640376). Most of computations were performed
on a 600 MHz Alpha PC164 workstation with a variant of Linux Alpha OS.

\newpage

\end{document}